\documentclass[sigconf,authorversion]{acmart}
\AtBeginDocument{%
  }

\copyrightyear{2026}
\acmYear{2026}
\setcopyright{none}
\acmConference[IWOCL '26]{International Workshop on OpenCL and SYCL}{May 06--08, 2026}{Heilbronn, Germany}
\acmBooktitle{International Workshop on OpenCL and SYCL (IWOCL '26), May 06--08, 2026, Heilbronn, Germany}
\acmDOI{10.1145/3811257.3811260}

\usepackage{algorithmic}
\usepackage{cleveref}
\usepackage{algorithm}
\usepackage{subcaption}
\usepackage{siunitx}
\usepackage{makecell}

\begin{document}

\title{Comparing the Performance of Heterogeneous Conjugate Gradient and Cholesky Solvers on Various Hardware Using SYCL}

\author{Tim Thüring}
\affiliation{%
  \institution{University of Stuttgart}
  \city{Stuttgart}
  \country{Germany}
}
\email{Tim.Thuering@ipvs.uni-stuttgart.de}
\orcid{0009-0006-1762-8429}

\author{Alexander Strack}
\affiliation{%
  \institution{University of Stuttgart}
  \city{Stuttgart}
  \country{Germany}
}
\email{Alexander.Strack@ipvs.uni-stuttgart.de}
\orcid{0000-0002-9939-9044}

\author{Dirk Pflüger}
\affiliation{%
  \institution{University of Stuttgart}
  \city{Stuttgart}
  \country{Germany}
}
\email{Dirk.Pflueger@ipvs.uni-stuttgart.de} 
\orcid{0000-0002-4360-0212}

\renewcommand{\shortauthors}{Thüring, Strack, Pflüger}

\begin{abstract}
Many important real-world applications, such as System Identification with Gaussian Processes, involve solving linear systems with symmetric positive-definite matrices.
The iterative CG method and direct solvers based on the Cholesky decomposition are two popular methods that can be applied in this case. 
Since often very large systems have to be solved when dealing with such real-world scenarios, GPUs are commonly used to accelerate the computations.
However, homogeneous approaches that only leverage the GPU in the system do not take full advantage of the often powerful CPUs located in modern HPC systems.
In this work, we present multi-vendor, heterogeneous implementations of the CG method and the Cholesky decomposition that leverage the CPU and GPU of a heterogeneous system simultaneously using SYCL. Furthermore, we compare their runtime behavior to traditional, homogeneous approaches.
The results show that for large matrices, our heterogeneous implementation is up to 32 percent faster for the CG method and up to 29 percent faster for the Cholesky decomposition compared to the corresponding GPU-only implementations.
In addition, for large matrices, our heterogeneous implementation of the Cholesky decomposition can achieve at least 12 percent faster runtimes across several systems with GPUs from NVIDIA, AMD, and Intel.
\end{abstract}

\ccsdesc[500]{Computing methodologies~Parallel programming languages}
\ccsdesc[300]{Mathematics of computing~Computations on matrices}

\keywords{Heterogeneous Computing, SYCL, Conjugate Gradient, Cholesky, Blocked Algorithms, Performance Comparison}


\maketitle

\section{Introduction}
\label{chap:intro}

Linear systems of the form $\boldsymbol{A}\cdot \boldsymbol{x} = \boldsymbol{b}$ involving a symmetric positive-definite (SPD) matrix $\boldsymbol{A}$ are fundamental building blocks of numerous important real-world applications. 
From the finite element method used in structural analysis~\cite{Chakraborty2002} to electromagnetic transient simulations~\cite{Maguire2011} and System Identification with Gaussian Processes (GPs)~\cite{Kocijan2016}, the solution of linear systems of equations is a core problem in science and engineering.
In this work, we consider the latter application to GPs, where direct solvers such as the Cholesky decomposition are commonly used, but also iterative solvers that yield an approximation to the solution, such as the Conjugate Gradient (CG) method~\cite{Hestenes1952} can be applied.

The linear systems can easily become very large, involving matrices with billions of entries. 
As a result, the computational effort to solve such systems dramatically increases, and GPUs are often used to accelerate the solving process.
With classic homogeneous approaches, the CPU launches a kernel on the GPU and thereby offloads the complete calculation to the GPU. 
The drawback of this approach is that during the time the GPU is performing the computation, the CPU waits for the GPU to finish and does not contribute to the solution.
However, many modern systems that are used for GPU computing feature powerful CPUs with high core counts and vectorization capabilities.
These computational resources are not used efficiently in classic homogeneous GPU-only approaches.
Heterogeneous computing targets this problem by aiming to leverage all available computational resources simultaneously to achieve a lower overall runtime.

Nevertheless, this comes with the inherent challenge of programmatically addressing the heterogeneous hardware. 
Conventionally, it requires different programming languages for this task, such as CUDA for NVIDIA GPUs or language extensions like OpenMP for multi-core CPUs.
As a result, libraries can end up with multiple instances of the same kernel written in different programming languages, which all require maintenance, optimization, and testing. 
In this work, we make use of SYCL to address this challenge of heterogeneous computing. 
By providing an abstraction layer, SYCL allows programming parallel kernels for various different architectures, including CPUs and GPUs of all major vendors, while relying on standard ISO C++ as a programming language.

There already exists related work that implements the CG algorithm or the Cholesky decomposition heterogeneously.
The \allowbreak pipelined, preconditioned variation of the CG method was implemented heterogeneously by~\citet{Tiwari2021} using a combination of CUDA and OpenMP.
A further heterogeneous implementation of the pipelined CG method was implemented by~\citet{Lang2013} using the same programming language combination as the previous authors. 
\citet{Nedozhogin2022} present a heterogeneous and distributed implementation of the preconditioned CG method and the pipelined CG method using MPI in addition to CUDA and OpenMP.

A heterogeneous implementation of the Cholesky decomposition was performed by~\citet{Ltaief2010} leveraging BLAS kernels of the MAGMA library ~\cite{Tomov2009} as building blocks in combination with task scheduling using the PLASMA library~\cite{Dongarra2019}, where they extended the static scheduler to support heterogeneous computing.
The MAGMA library also has ongoing efforts to implement a SYCL backend \cite{Abdelfattah2024}.
Further heterogeneous Cholesky decomposition implementations include the works by~\citet{Song2012, Song2015} and~\citet{Alonso2012}.

Furthermore, homogeneous implementations of the CG algorithm and the Cholesky decomposition exist that make use of SYCL. 
\citet{Cali2021} compared their SYCL implementation of the CG algorithm using DPC++ on CPUs, GPUs and FPGAs. 
The work by~\citet{Baratta2022} compares a SYCL implementation of the matrix-free CG variant against reference libraries using CPUs and GPUs.
Another CG implementation that makes use of SYCL is included in the PLSSVM library by~\citet{VanCraen2022}.
PLSSVM mainly focuses on homogeneous computing; however, in theory, heterogeneous computing could be achieved using a combination of SYCL and MPI.
With the \texttt{potrf} routine, the Cholesky decomposition is exposed via the SYCL API of the oneAPI Math Library (oneMath)~\cite{UXL} developed by the Unified Acceleration (UXL) Foundation. 
Even though this enables the usage of the Cholesky decomposition via a SYCL API, the backend is not necessarily SYCL code, as vendor-specific libraries can be used in the background.

Generally, most existing heterogeneous implementations of the CG method and the Cholesky decomposition rely on a combination of several programming languages or language extensions to achieve simultaneous execution on CPU and GPU. 
Therefore, we have developed heterogeneous implementations of the CG method and the Cholesky decomposition using SYCL from scratch~\cite{thueringtim}.
This allows for a fair comparison of the heterogeneous and homogeneous implementations on various systems featuring GPUs from NVIDIA, AMD, and Intel.
To our knowledge, such a comparison of CG and Cholesky solvers for solving Gaussian Process covariance matrices in a heterogeneous setting using solely SYCL has not been performed before.
In summary, this work makes the following contributions:
\begin{itemize}
    \item Novel, native SYCL implementations of memory-efficient, block-based CG and Cholesky solvers targeting both homogeneous and heterogeneous computing,
    \item Performance evaluation of the heterogeneous and homogeneous solvers on various hardware, including GPUs from NVIDIA, AMD, and Intel,
    \item Comparison of AdaptiveCpp and the Intel oneAPI DPC++/C++ compiler icpx regarding the CPU, GPU, and heterogeneous performance of both solvers.
\end{itemize}
This paper is structured as follows: first, the CG method and the Cholesky decomposition are introduced in \Cref{chap:fund}. Subsequently, we explain how we implemented both methods heterogeneously on the CPU and GPU in \Cref{chap:impl} before we discuss the results of the performance analysis of the heterogeneous approach in \Cref{chap:res}.
Last, in \Cref{chap:conl}, we conclude our findings and provide an outlook for future work.

\section{Fundamentals}
\label{chap:fund}
This chapter provides all relevant mathematical fundamentals regarding the two solvers for linear systems considered in this work: the CG method and the Cholesky decomposition.

\begin{algorithm}[htb!]
\begin{minipage}{.23\textwidth}
	\begin{algorithmic}[1]
	\STATE $\boldsymbol{s} = \boldsymbol{r} = \boldsymbol{b} - \boldsymbol{A}\boldsymbol{x}$
	\STATE $u = u_0 = \boldsymbol{r}^T\boldsymbol{r}$
	\WHILE{$u > \epsilon^2 u_0$}
		\STATE $\boldsymbol{t} = \boldsymbol{A}\boldsymbol{s}$
		\STATE $\alpha = \frac{u}{\boldsymbol{s}^T\boldsymbol{t}}$
		\STATE $\boldsymbol{x} = \boldsymbol{x} + \alpha \boldsymbol{s}$
		\STATE $\boldsymbol{r} = \boldsymbol{r} - \alpha \boldsymbol{t}$
		\STATE $v = u$
		\STATE $u = \boldsymbol{r}^T\boldsymbol{r}$
		\STATE $\beta = \frac{u}{v}$
		\STATE $\boldsymbol{s} = \boldsymbol{r} + \beta \boldsymbol{s}$
	\ENDWHILE
	\end{algorithmic}
\end{minipage}
\begin{minipage}{.23\textwidth}
    \begin{algorithmic}[1]
        \FOR{j = 0...N-1}
            \STATE $\boldsymbol{A}_{jj} = \texttt{Cholesky}(\boldsymbol{A}_{jj})$
            \FOR{i = j+1...N-1}
                \STATE $\boldsymbol{A}_{ij} = \boldsymbol{A}_{ij} \cdot \boldsymbol{A}_{jj}^{-\top}$
            \ENDFOR
            \FOR{i = j+1...N-1}
                \STATE $\boldsymbol{A}_{ii}  \mathrel{-}= \boldsymbol{A}_{ij} \cdot \boldsymbol{A}_{ij}^\top$
                \FOR{k = j+1...i-1}
                    \STATE $\boldsymbol{A}_{ik} \mathrel{-}= \boldsymbol{A}_{ij} \cdot \boldsymbol{A}_{kj}^\top$
                \ENDFOR
            \ENDFOR
        \ENDFOR
	\end{algorithmic}
\end{minipage}

\caption{The CG algorithm (left) and the in-place right-looking blocked Cholesky decomposition (right). CG code modified from~\citet{Shewchuk1994} page 50, Section B2. 
Cholesky code modified from~\citet{Dorris2016} Fig. 1 and~\citet{VanDeGeijn2007} Fig. 6.6.}
\label{alg:cg_chol}
\end{algorithm}

\subsection{The Conjugate Gradient method}
\label{sec:fund_cg}
The CG method by~\citet{Hestenes1952} is an iterative solver for linear systems of the form $\boldsymbol{A}\boldsymbol{x} = \boldsymbol{b}$ that approximates the solution $\boldsymbol{x}$.
Specifically, the method requires the matrix $\boldsymbol{A}$ of the linear system to be SPD. 

The left column of \Cref{alg:cg_chol} shows pseudocode for the CG algorithm as in \allowbreak~\citet{Shewchuk1994}.
In every iteration, the CG algorithm performs the same operations from lines four to eleven and iteratively refines the solution until it is below the residual tolerance that can be adjusted via the $\epsilon$ value. 
Only one matrix-vector product is required per iteration.
The remaining operations solely involve vectors or are purely scalar.
To compensate for rounding errors that stem from the update of the residual in line seven, the actual residual $ \boldsymbol{b} - \boldsymbol{A}\boldsymbol{x}$ has to be recomputed from scratch every few iterations.
Thus, a second matrix vector product is required in these iterations.

\subsection{The Cholesky decomposition}
\label{sec:fund_chol}
As the CG algorithm, the Cholesky decomposition can only be applied to linear systems $\boldsymbol{A}\boldsymbol{x} = \boldsymbol{b}$ where the matrix $\boldsymbol{A}$ is SPD.
However, in contrast to the CG method, it is a direct solver.
When applying the Cholesky decomposition to $\boldsymbol{A}$, the matrix is factored into a lower triangular matrix $\boldsymbol{L}$ with the property that $\boldsymbol{L}\boldsymbol{L}^\top = \boldsymbol{A}$.
When such a lower triangular matrix is computed, the resulting system $\boldsymbol{L}\boldsymbol{L}^\top\boldsymbol{x} = \boldsymbol{b}$ can be solved efficiently using a forward and back substitution.



In this work, we consider the blocked, right-looking variant of the Cholesky decomposition, as explained by e.g.,~\citet{Dorris2016} and~\citet{VanDeGeijn2007}.
The corresponding pseudocode is shown in the right column in \Cref{alg:cg_chol}.
In this version of the algorithm, the matrix is partitioned into square blocks of equal size.
The algorithm works in-place and transforms the lower triangular part of the matrix $\boldsymbol{A}$ into $\boldsymbol{L}$.
For this purpose, the algorithm iterates through the blocked columns of the matrix from left to right.
In each of these column iterations, the same sequence of operations is performed on the blocks of the matrix.
First, in \textit{Step 1}, the diagonal block of the current column is factored into a lower triangular matrix using a standard Cholesky decomposition in line two. 
Subsequently, in \textit{Step 2}, the remaining column below is updated by solving the matrix equation in line four, which results in solving a triangular system with multiple right-hand sides for each block in the column below the diagonal.
Finally,  in \textit{Step 3}, all blocks to the right of the current column are updated with matrix-matrix multiplications in lines seven and nine.

\section{Implementation}
\label{chap:impl}
This chapter contains details about our heterogeneous implementation of the CG method and the Cholesky decomposition.
To achieve heterogeneous execution on the CPU and GPU simultaneously, the implementation has to be capable of targeting both architectures.
Our implementation makes use of SYCL~\cite{KhronosGroup}, which allows targeting different architectures for parallel execution while relying on standard ISO C++ for programming. 
The SYCL standard is developed by the Khronos Group\footnote{\url{https://www.khronos.org} (visited on 01/06/2026)}.
Multiple different implementations of the SYCL standard exist. 
In this work we make use of the AdaptiveCpp\footnote{\url{https://github.com/AdaptiveCpp/AdaptiveCpp} (visited on 01/06/2026)}~\cite{Alpay2020} SYCL implementation and the Intel oneAPI DPC++/C++ compiler\footnote{\url{https://www.intel.com/content/www/us/en/developer/tools/oneapi/dpc-compiler.html} (visited on 01/06/2026)} (icpx). 

As explained in \Cref{chap:fund}, a common property of the CG method and the Cholesky decomposition is that both solvers require the matrix $A$ of the system to be SPD.
The symmetry of the matrices can be exploited, such that it is not necessary to store the complete matrix in memory.
Our implementation partitions the matrix into square blocks and only stores the lower-triangular and diagonal blocks. 
As a result, solely the diagonal blocks store some redundant data.

All performance-critical kernels are implemented and parallelized from scratch without the use of vendor-specific libraries. 
This decision was made to ensure that the performance measurements are not biased through unequal degrees of optimization between different vendor libraries.
In the following sections, we explain how we implemented the CG algorithm and the Cholesky decomposition heterogeneously on the CPU and GPU.

\subsection{CG implementation}
\label{sec:cg_impl}
The concept of how the workload of the CG algorithm is distributed between the CPU and GPU is based on the approach of~\citet{Tiwari2021}, who implemented the pipelined, preconditioned CG with their \textit{"Hybrid-PIPECG-3"} algorithm.
The runtime of the CG algorithm is dominated by the matrix-vector product, which is memory-bound.
In the hybrid approach developed by the authors, the matrix-vector product and all other vector operations are distributed between the CPU and GPU and calculated cooperatively.
In this work, we adapted the approach to work with the data layout for symmetric matrices and the classic version of the CG algorithm that is considered in this work.

All matrix-vector and vector-vector operations that comprise the CG algorithm are computed cooperatively by the CPU and the GPU. 
The matrix is split horizontally between two block rows of the blocked matrix data layout.
All vectors are split at the same row as the matrix. 
The height at which the matrix is split can be chosen such that the workload between the CPU and GPU is balanced.
The CPU computes the lower part of the result vector of the matrix-vector product and all lower parts of the result vectors emerging from the remaining operations.
The GPU performs the analogous computations involving the upper parts of the vectors.

With this approach, only a small amount of communication is needed between the CPU and the GPU.
For the vector-vector operations that are scalar products, two partial sums are computed on the CPU and GPU separately.
To obtain the final result value, the result of the partial sum computed by the GPU has to be explicitly copied to the CPU.
This situation occurs twice in the CG algorithm, i.e., in lines five and nine in~\Cref{alg:cg_chol}.
Furthermore, communication of a sub-vector is required once in each iteration. 
When the vector $\boldsymbol{s}$ is updated at the end of each iteration by the CPU and GPU in line eleven, the result is distributed in the two different memory spaces of the two kinds of hardware. 
However, the complete vector $\boldsymbol{s}$ is required by the CPU and the GPU at the beginning of the next iteration in line four, and thus, communication of the sub-vectors is required.
For the remaining vector operations in the CG algorithm, this is not a problem since these operations only require previously computed values located in the sub-vector that is assigned to the corresponding type of hardware.
Thus, all required values are already located in the correct memory space. 
As described in~\Cref{sec:fund_cg}, the real residual is sometimes recomputed from scratch instead of the update in line seven. 
In this case, a second matrix-vector product and, thus, also an additional vector communication is necessary in the affected iterations.

The SYCL kernels for the vector-vector and matrix-vector operations are implemented from scratch and optimized for CPUs and GPUs individually.
The GPU-scalar product kernel is based on the fourth version of the implementation by~\citet{Harris}.
Our matrix-vector product kernel was adapted from the implementation performed by~\citet{Nath2011}.

\subsection{Cholesky implementation}
\label{sec:chol_impl}
Our implementation of the Cholesky decomposition follows the blocked right-looking approach.
The blocks of the matrix are updated using different operations, such as a single block Cholesky decomposition, a triangular matrix solve, and both symmetric and classical matrix-matrix multiplications.
The runtime is dominated by the compute-bound matrix-matrix operations.
The algorithm modifies the lower blocked triangular matrix and transforms it into the desired result matrix $\boldsymbol{L}$. 
In the right-looking algorithm, one column of the matrix is processed after another from left to right, applying the three steps defined in~\Cref{sec:fund_chol}.
For each column iteration, a horizontal border is defined that splits the matrix blocks between CPU and GPU.
All blocks located above the border are processed by the CPU, while all blocks below are updated by the GPU.
As a result, the communication of blocks between the CPU and GPU is necessary. 
In each column iteration, the following blocks  have to be sent to the GPU, as their values are required for the next processing steps: the current diagonal block, and the blocks in the sub-column below processed by the CPU in \textit{Step 1} and \textit{Step 2}.

One property of the right-looking Cholesky decomposition is that the triangular sub-matrix, modified by the algorithm in each column iteration, gets smaller as the algorithm proceeds.
This has implications for the heterogeneous implementation since, in contrast to the CG method, the workload is not constant over time.
The matrix is split horizontally between two block rows to distribute the workload between the CPU and GPU.
However, if this split remains constant throughout the execution, the CPU runs out of work.
Since the sub-matrix that is updated becomes smaller with every column iteration, as the leftmost blocks are processed completely, the sub-matrix above the split gets smaller.
To avoid this imbalance in blocks assigned to the CPU and GPU, the row where the matrix is divided between the CPU and GPU has to be shifted down every few iterations to keep the proportion of assigned blocks roughly constant over time.
However, such a shift of the border between the CPU and GPU requires the communication of a whole block row.

As for the CG Algorithm, the individual SYCL kernels required for the Cholesky decomposition are implemented from scratch and optimized separately for CPUs and GPUs.
The GPU kernel implementation of the computationally most expensive part, the update of all matrix blocks below the diagonal in \textit{Step 3}, is based on a scaled-down version of the implementation by \citet{Tan2011}.
The GPU kernel that updates the blocks on the diagonal using symmetric matrix-matrix multiplications is adapted from the implementation in \cite{Rauber2023}, page 447, Figure 7.13.

\section{Results}
\label{chap:res}

This chapter presents the results of our comparison of the runtime behavior of the heterogeneous and homogeneous implementations for the CG algorithm and the Cholesky decomposition.
After our experimental setup is outlined, the two algorithms are first considered separately.
For each algorithm, the optimal parameter configuration on the CPU and GPU is determined first.
Subsequently, the heterogeneous and homogeneous implementations are compared.
Next, a comparison between AdaptiveCpp and the Intel oneAPI DPC++/C++ compiler (icpx) is conducted for the respective algorithm.
Finally, the CG algorithm and the Cholesky algorithm are compared against each other on various multi-vendor hardware.

\subsection{Experimental Setup}
\label{sec:results_setup}

The hardware we use to evaluate the heterogeneous implementations is listed in \Cref{tab:systems}. 
First, we focus on \textit{System 1} and \textit{System 2}, which feature an identical dual-socket 48-core AMD EPYC CPU. 
In addition, \textit{System 1} has an NVIDIA A30 GPU installed, and \textit{System 2} features an AMD MI210 GPU.
Later, we extend the comparison to \textit{System 3} and \textit{System 4}, which are equipped with an identical 18-core Intel CPU.
Furthermore, an Intel Arc B580 GPU is installed in \textit{System 3}, and an NVIDIA RTX 3080 is installed in \textit{System 4}.
The stated FP64 peak performance values for the CPUs are calculated manually using the CPU base frequency and the available AVX units.

\begin{table}[htb!]
	\centering
    \small
	\begin{tabular}{|c||cc|cc|}
		\hline
		& \multicolumn{1}{c|}{\textbf{System 1}}   &\textbf{ System 2}           & \multicolumn{1}{c|}{\textbf{System 3}}            & \textbf{System 4 }       \\ \hline\hline
		\textbf{CPU}            & \multicolumn{2}{c|}{2x AMD EPYC 9274F}               & \multicolumn{2}{c|}{Intel Core i9-10980XE}                 \\ \hline
		\textbf{\makecell{Cores/ \\ Threads}}  & \multicolumn{2}{c|}{\makecell{48/96 \\ (combined)}}                           & \multicolumn{2}{c|}{18/36}                                 \\ \hline
		\makecell{\textbf{CPU FP64} \\ \textbf{max FLOPS}} & \multicolumn{2}{c|}{\makecell{3.1104 TFLOPS \\  (combined)}}                                & \multicolumn{2}{c|}{1.728 TFLOPS}                                      \\ \hline
		\textbf{RAM}            & \multicolumn{2}{c|}{384GB DDR5}                      & \multicolumn{2}{c|}{64GB DDR4}                             \\ \hline\hline
		\textbf{GPU}            & \multicolumn{1}{c|}{\makecell{NVIDIA \\ A30}} & \makecell{AMD \\ MI210} & \multicolumn{1}{c|}{\makecell{Intel \\ B580}} & \makecell{NVIDIA \\ 3080} \\ \hline
		\makecell{\textbf{GPU FP64}  \\\textbf{max FLOPS}} & \multicolumn{1}{c|}{\makecell{5.2 \\ TFLOPS}}           &          \makecell{22.6 \\ TFLOPS }        & \multicolumn{1}{c|}{\makecell{N/A}}                    &        \makecell{ 0.466 \\ TFLOPS }      \\ \hline
		\textbf{\makecell{GPU \\ memory}}     & \multicolumn{1}{c|}{\makecell{24GB \\ HBM2 \\(933 GB/s)}}       & \makecell{ 64GB  \\ HBM2e \\(1.6 TB/s)}              & \multicolumn{1}{c|}{\makecell{12GB \\ GDDR6 \\ (456 GB/s)}}                   & \makecell{10GB \\ GDDR6X \\ (760 GB/s)}           \\ \hline
	\end{tabular}
	\caption
    {Overview of the four test systems used in our evaluation.
    The hardware specifications of the systems were retrieved from the following sources: 
		AMD EPYC: \cite{AMD2023}, \cite{Bhargava2024}; 
        NVIDIA A30: \cite{NVIDIA2022}; 
        AMD MI210: \cite{AMD2022}; 
        Intel i9: \cite{Intelb}; 
        Intel B580: \cite{Intela}; 
        NVIDIA 3080: \cite{NVIDIA2021}.		
	}
\label{tab:systems}
\end{table}

If not stated otherwise, AdaptiveCpp is used.
We use AdaptiveCpp v25.02.0, which is built against LLVM 19.1.0.
For the AdaptiveCpp measurements on the GPUs, we use CUDA 12.2.2 on \textit{System 1} and CUDA 12.4.1 on \textit{System 4}, ROCM 6.4.0 on \textit{System 2}, and oneAPI 2025.1 on \textit{System 3} for the respective backends.
When comparing AdaptiveCpp against the Intel oneAPI DPC++/C++ compiler (icpx), we use icpx version 2025.1.1 based on LLVM 20.0.0.
For the icpx measurements on GPUs, we use CUDA 12.6.3 and ROCM 6.4.0 for the respective backends.
We use the AdaptiveCpp \textit{cuda}/\textit{hip} and \textit{omp.accelerated} compilation flows for systems with NVIDIA and AMD GPUs and the AdaptiveCpp \textit{generic} compilation flow for Intel GPUs. 
Using the \textit{generic} compilation flow with an otherwise identical configuration on NVIDIA and AMD GPUs resulted only in marginally faster GPU runtimes, which lay within one percent of the \textit{cuda}/\textit{hip} compilation flow runtimes.
The intel icpx compiler uses \texttt{-fp-model=fast} as a default setting, whereas AdaptiveCpp does not use similar optimizations. However, when using \texttt{-fp-model=precise}, it only slightly increases the runtimes in our setup, e.g., by well below one percent on the CPU, thus, we use the default settings for icpx.

As input data, we generated kernel matrices based on simulated data of a mass-spring-damper system by \citet{Helmann2025a} available at \cite{Helmann2025}. 
The kernel matrices can be used for behavior prediction of the system using GP as described by \citet{Kocijan2016}. 
In such a case, a linear system that involves the SPD kernel matrix has to be solved.

All measurements are averaged over at least ten runs and are performed using FP64 double precision.
The CG $\epsilon$-value that affects the termination condition is set to $10^{-6}$.
Due to rounding errors, the exact iteration in which the CG algorithm terminates varies slightly between different kernels. To reduce the impact on runtime variation, we limited the maximum number of iterations to 60, 70, 75, 80, and 95 iterations, depending on the size of the dataset.
In the last section, when we compare the Cholesky algorithm to the CG algorithm, the number of iterations is not limited such that both methods solve the same problem.
For the Cholesky algorithm, the forward and backward substitutions are included.
Otherwise, only the Cholesky decomposition is considered.


\subsection{Heterogeneous CG Evaluation}
\label{sec:results_CG_hetperf}
In this section, our heterogeneous implementation of the CG method is evaluated on \textit{System 1} and \textit{System 2} using AdaptiveCpp. 
First, we determine the optimal values for the parameters that influence the runtime behavior of the algorithm.
Subsequently, we analyze different workload distributions between the CPU and GPU before comparing the heterogeneous approach to the homogeneous CPU-only and GPU-only implementations.

\subsubsection{Optimal CPU and GPU Configuration for the CG Algorithm}
\label{sec:results_CG_opt}

There are several parameters that can be used to configure the CG algorithm on CPUs and GPUs.
Since AdaptiveCpp uses OpenMP as a backend for the parallelization on CPUs, the OpenMP configuration, such as the number of threads and the usage of thread binding, influences the runtime on CPUs.
Furthermore, the block size of the blocked matrix data structure influences the work-group sizes of the underlying SYCL kernels and thus has an impact on the performance.

To determine the optimal OpenMP configuration for the CG algorithm, we evaluated different configurations on the 48-core dual-socket CPU of \textit{System 1} using a matrix with side length $65536$.
We compared the performance of the homogeneous CG algorithm on the CPU using 48 and 96 threads, which corresponds to disabling/enabling simultaneous multi-threading (SMT) on the CPU, and with thread binding enabled or disabled.
Furthermore, since it is unclear if the CPU's Advanced Vector Extension (AVX) units provide an advantage for the memory-bound CG algorithm, all configurations are compared with and without AVX.
We observed that using AVX does not result in lower runtimes in most cases, and 48 threads provided a better performance than 96 threads most of the time.
For example, when using 48 threads and enabling binding by setting the \texttt{OMP\_PROC\_BIND} environment variable to \texttt{true}, the CG algorithm needs about \num{47.52297}\si{\second} with and \num{33.22680}\si{\second}  without AVX.
When increasing the thread count to 96 in the previous scenario, the runtimes increase to \num{52.81911}\si{\second} and \num{50.20831}\si{\second} respectively.
Furthermore, enabling thread binding lowered the variations in the runtime between different runs while providing a similar average performance.
Thus, we choose an OpenMP configuration that does not make use of SMT and enables binding for all further experiments on CPUs and AdaptiveCpp.

The next parameter that can be varied for the CG algorithm is the block size of the underlying matrix data structure.
Since it influences the work-group size of the CPU and GPU SYCL kernels, it has a crucial impact on the performance.
For all devices that are used in this work, the optimal block size has been determined by comparing the performance for all powers of two from 16 to 1024 as a value for the side length of the square blocks.
The results show that on the AMD CPU of \textit{System 1} and \textit{System 2}, a block size of 32 performs best with a runtime of \num{33.1677009451}\si{\second}. The runtime can vary significantly when selecting a different block size value. For example, when choosing a value of 1024, the runtime increases to \num{139.3235282444}\si{\second}, highlighting the importance of tuning this variable for the respective hardware.
For the Intel CPU in \textit{System 3} and \textit{System 4}, a value of 16 performed best.
On the NVIDIA A30 GPU, a value of 64 resulted in the lowest runtime.
The AMD MI210 and NVIDIA RTX 3080 GPUs performed best for 32.
Finally, the Intel GPU in \textit{System 3} achieved the best performance for blocks of size $256 \times 256$.



We can observe that the optimal block size varies widely between different devices.
Also for GPUs, the runtime differences between optimal and non-optimal block sizes can be non-negligible.
For example, on the Intel Arc B580 GPU, choosing a block size of 32 instead of the optimal 256 results in \num{1.8692586298545673} times slower runtimes.



\subsubsection{Optimal Heterogeneous Workload Distribution}
\label{sec:results_CG_hetperf_dist}

The workload distribution between the CPU and GPU is a crucial variable when performing a computation heterogeneously.
To leverage the full potential of the two devices during heterogeneous execution, it is desirable that the CPU and GPU finish their workload at the same time.
Otherwise, if one device finishes earlier than the other, the faster device has to wait for the slower device, which results in a non-optimal utilization of the hardware.
Since the CPU and GPU generally have vastly different performance, finding the optimal workload distribution for a given system is a non-trivial task.

\begin{figure}[htb]
    \centering
    \includegraphics[width=\columnwidth]{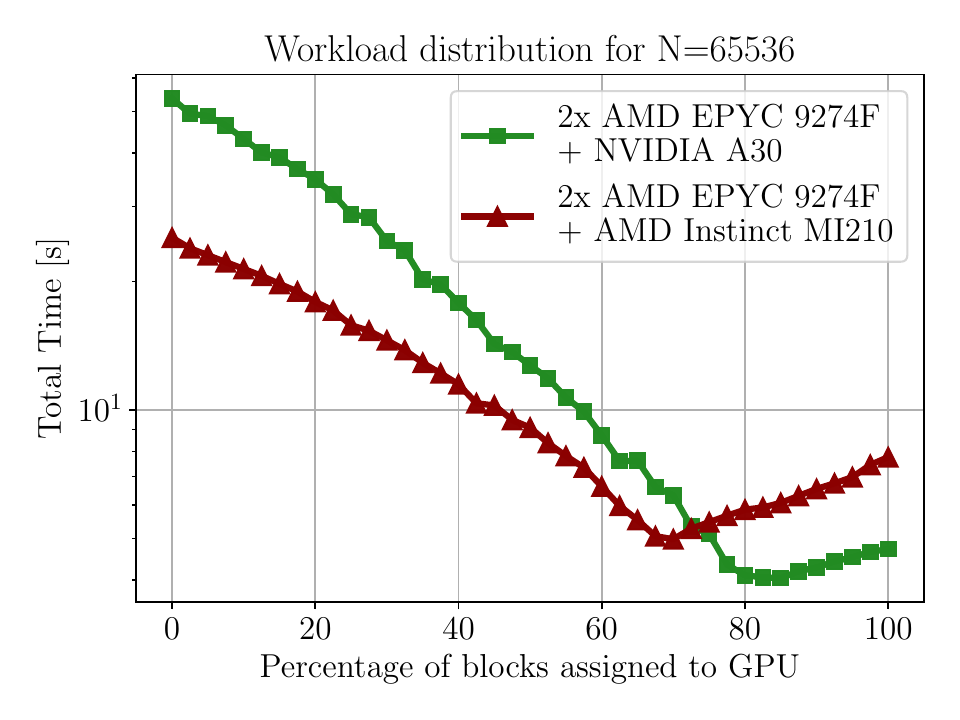}
    \caption{Different workload distributions for the heterogeneous CG algorithm on \textit{System 1} and \textit{System 2}.
    On \textit{System 1}, the workload distribution is optimal when 85\% of the matrix blocks are assigned to the GPU, whereas this optimum corresponds to 70\% on \textit{System 2}.}
    \label{fig:cg_het_split_S1_S2}
\end{figure}

\Cref{fig:cg_het_split_S1_S2} compares the performance of the heterogeneous CG implementation with different workload distributions on \textit{System 1} and \textit{System 2} to find the optimal assignment for the respective system.
Both systems are equipped with an identical CPU and different GPUs, which makes them suitable for a direct comparison of the workload distribution.
The x-axis denotes the percentage of blocks that are assigned to the GPU, and the y-axis corresponds to the runtime of the heterogeneous CG algorithm with the given workload distribution.
The runtime does not include the memory transfer times between the CPU and GPU at the beginning and the end.
The matrix size chosen for this experiment corresponds to $65536 \times 65536$.
For both systems, an optimal heterogeneous workload distribution can be found. 
On \textit{System 1}, the lowest runtime can be achieved when assigning \num{85}\si{\percent} of the workload to the NVIDIA A30 GPU.
When considering the results obtained on \textit{System 2}, we can observe an optimal assignment when \num{70}\si{\percent} of the work is performed by the AMD MI210 GPU.
This observation contrasts with what we would expect when solely considering the theoretical performance of the two GPUs. 
As the CPU in both systems is identical and the AMD GPU offers a much higher theoretical memory bandwidth, one would expect that more work is assigned to the AMD GPU in \textit{System 2} than to the NVIDIA GPU in \textit{System 1}.
However, this is not the case.
The NVIDIA GPU in \textit{System 1} has a higher proportion of the work assigned to it.
Additionally, the overall runtime of \textit{System 1} for the heterogeneous CG algorithm is lower than on \textit{System 2}.
Thus, the peak memory bandwidth alone is not sufficient to explain the behavior, and the capabilities of the complete memory hierarchy have to be considered.
Here, the NVIDIA A30 GPU has an advantage since with 24MB, its L2 cache is much larger than that of the AMD MI210 GPU, which is only 8MB (values retrieved via the CUDA/HIP API).
However, we have to note that, as we will show in \Cref{sec:results_CG_app_icpx}, AdaptiveCpp does perform worse than icpx on the AMD GPU. 
Nevertheless, icpx is also not able to achieve faster runtimes with the AMD MI210 than with the NVIDIA A30.

Another observation is that \textit{System 2} performs much better when the heterogeneous CG algorithm is CPU-bound.
Since both systems are equipped with identical CPUs, identical runtimes would be expected.
However, the block size of the matrix was not chosen equally on the two systems. 
Since most of the work is performed on the GPU in both cases, the block size that is optimal for the GPU is chosen for the heterogeneous execution.
Since these block sizes do not perform equally well on the CPU, the performance for the CPU-bound case differs on the two systems.

\subsubsection{Comparison of the Heterogeneous and Homogeneous CG Performance}
\label{sec:results_CG_hetperf_rt}

This section compares the heterogeneous and homogeneous performance of the CG implementation for different matrix sizes on \textit{System 1} and \textit{System 2}.
Similar to the last section, the optimal heterogeneous workload distributions have been determined for the remaining matrix sizes on both systems. For the largest three matrices, the optimal percentages of blocks assigned to the GPU are relatively close to each other as they fall into a range between \num{82.5}\si{\percent} and \num{87.5}\si{\percent} on \textit{System 1} and between \num{65}\si{\percent} and \num{70}\si{\percent} on on \textit{System 2}.

\begin{figure}[htb]
    \centering
    \includegraphics[width=\columnwidth]{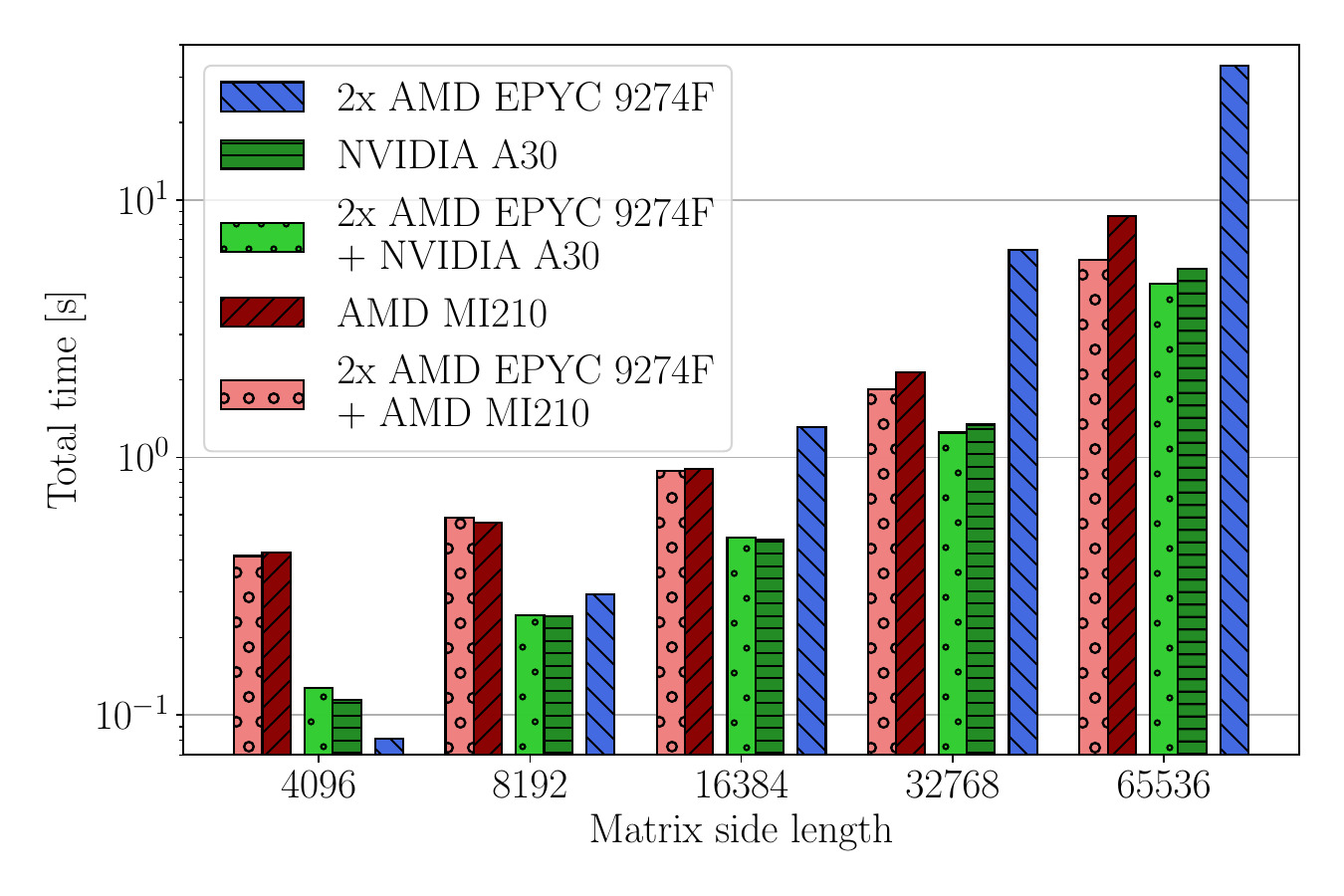}
    \caption{Heterogeneous and homogeneous runtime comparison for the CG algorithm on all hardware of \textit{System 1} and \textit{System 2}. On both systems, the heterogeneous approach results in a better performance for large matrices.}
    \label{fig:cg_perf_S1_S2}
\end{figure}

\Cref{fig:cg_perf_S1_S2} shows the results of the comparison.
The x-axis lists the different matrix sizes, and the y-axis corresponds to the runtime, which includes the memory transfer times between the CPU and GPU at the beginning and the end.
The axes are logarithmically scaled.
For large matrices, the CPU-only implementation results in the slowest runtime. 
When solving a linear system with a matrix of size $65536 \times 65536$, the dual-socket AMD EPYC 9274F CPU of \textit{System 1} and \textit{System 2} takes \num{33.1690922267}\si{\second}.
In contrast, the corresponding homogeneous GPU-only execution takes only \num{5.389919885299999}\si{\second} on the NVIDIA A30 GPU and \num{8.6799486113}\si{\second} on the AMD MI210 GPU.
When performing the CG algorithm heterogeneously on the CPU and GPU, performance improvements can be observed for the two largest matrix sizes on both systems.
When considering the runtime for the largest matrix, \textit{System 1} can finish the computation of the CG algorithm in \num{4.7143232742}\si{\second}, which corresponds to a relative improvement of \num{12.534446252950127}\si{\percent} regarding the homogeneous runtime obtained on the NVIDIA A30.
\textit{System 2} needs about \num{5.82859367455}\si{\second} for the heterogeneous computation, which translates into a relative performance improvement of \num{32.849905736054247}\si{\percent} over the homogeneous runtime of the AMD MI210 GPU in the system.
Overall, the findings are consistent with the observations made in \Cref{sec:results_CG_hetperf_dist}.
The NVIDIA GPU seems to be much better suited for the memory-bound CG algorithm. 
This also affects the heterogeneous performance improvements and explains why the relative performance improvement is so much higher on \textit{System 2}.
Since the optimal CPU proportion of the heterogeneous execution is much higher on \textit{System 2} (30\si{\percent}  compared to just 15\si{\percent}), the relative improvement is also expected to be higher.
For smaller matrix sizes, the heterogeneous CG implementation is not able to achieve performance improvements in most cases.
However, for larger matrices, the improvements are quite considerable.

\subsection{Comparison of AdaptiveCpp and Intel icpx using the CG algorithm}
\label{sec:results_CG_app_icpx}
In our previous experiments, we have focused on the AdaptiveCpp SYCL implementation. 
However, other popular SYCL implementations like the Intel oneAPI DPC++/C++ Compiler icpx exist.
In this section, the heterogeneous and homogeneous performance of our CG implementations is compared using the two SYCL implementations.
The optimal block sizes and heterogeneous workload distributions have been determined using the Intel SYCL implementation in the same manner as for AdaptiveCpp. 
The CPU configuration for icpx corresponds to the default values since they already provide near-optimal performance.

\begin{figure}[htb]
    \centering
    \includegraphics[width=\columnwidth]{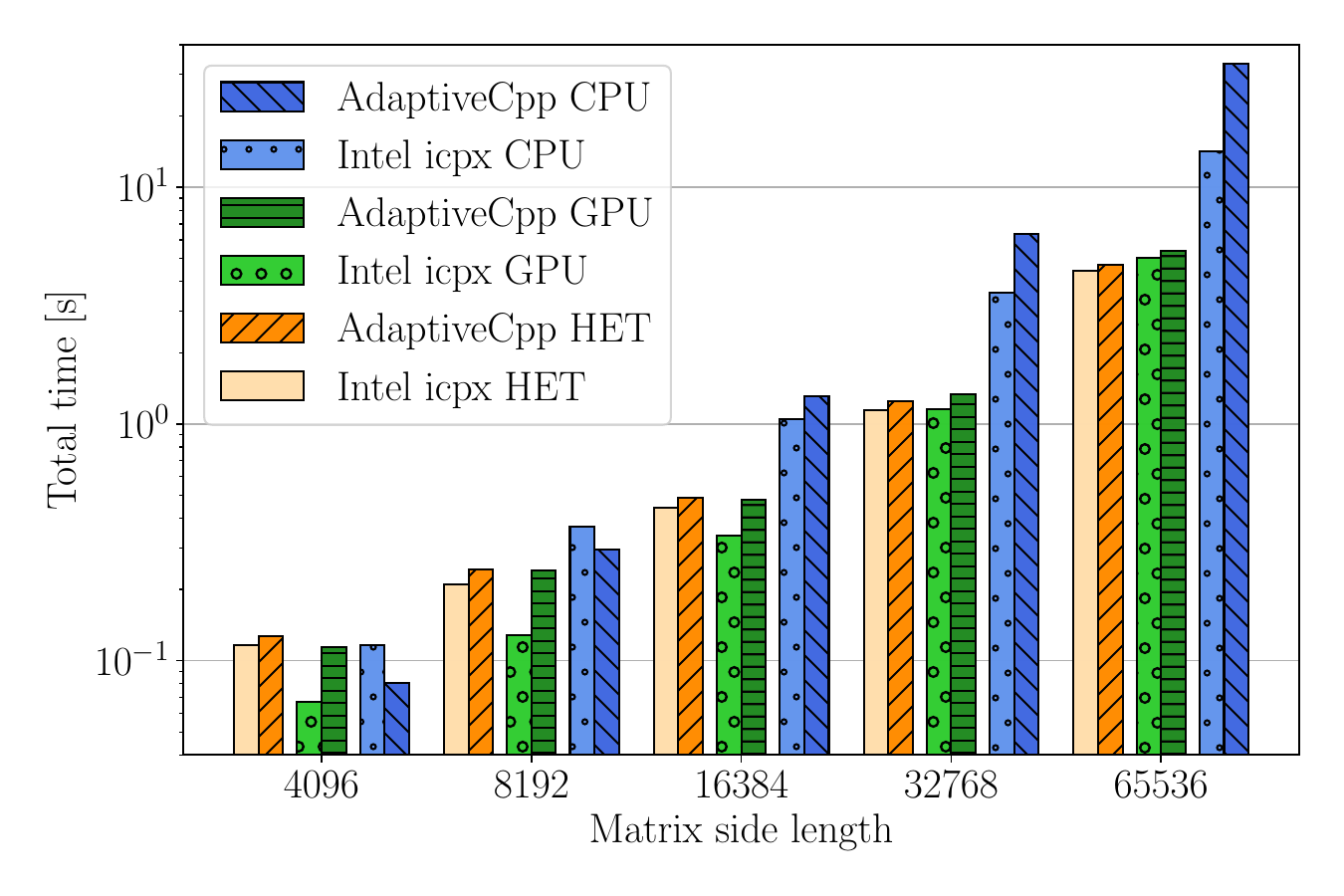}
    \caption{Comparison of the homogeneous and heterogeneous CG algorithm on \textit{System 1} with an NVIDIA A30 GPU using AdaptiveCpp and the Intel oneAPI DPC++/C++ compiler (icpx). In most cases, icpx results in faster runtimes.}
    \label{fig:cg_icpx_acpp_s1}
\end{figure}

The results of the comparison on \textit{System 1} featuring an NVIDIA A30 GPU are shown in \Cref{fig:cg_icpx_acpp_s1}.
When considering the CPU-only runtimes, we can observe that the Intel compiler results in faster runtimes for larger matrix sizes.
For example, the CG algorithm takes only \num{14.2128045084}\si{\second} to complete when the program is compiled with icpx, which is about \num{57.1504567828987}\si{\percent}  faster than the runtime of  \num{33.1690922267}\si{\second} when using AdaptiveCpp.
However, for smaller matrix sizes, AdaptiveCpp results in better performance.
When considering the GPU-only runtimes on the NVIDIA A30 GPU for the largest matrix, the CG algorithm finishes after \num{5.3899198853}\si{\second} when using AdaptiveCpp. 
The Intel icpx compiler is about \num{6.65951924218667}\si{\percent} faster with a runtime of \num{5.0309771334}\si{\second}.
A similar observation can be made for the heterogeneous case in which the CG algorithm takes about \num{4.4245347703}\si{\second} for the largest matrix with the icpx compiler, which is \num{6.14697989605248}\si{\percent} faster than the corresponding AdaptiveCpp runtime.

\begin{figure}[htb]
    \centering
    \includegraphics[width=\columnwidth]{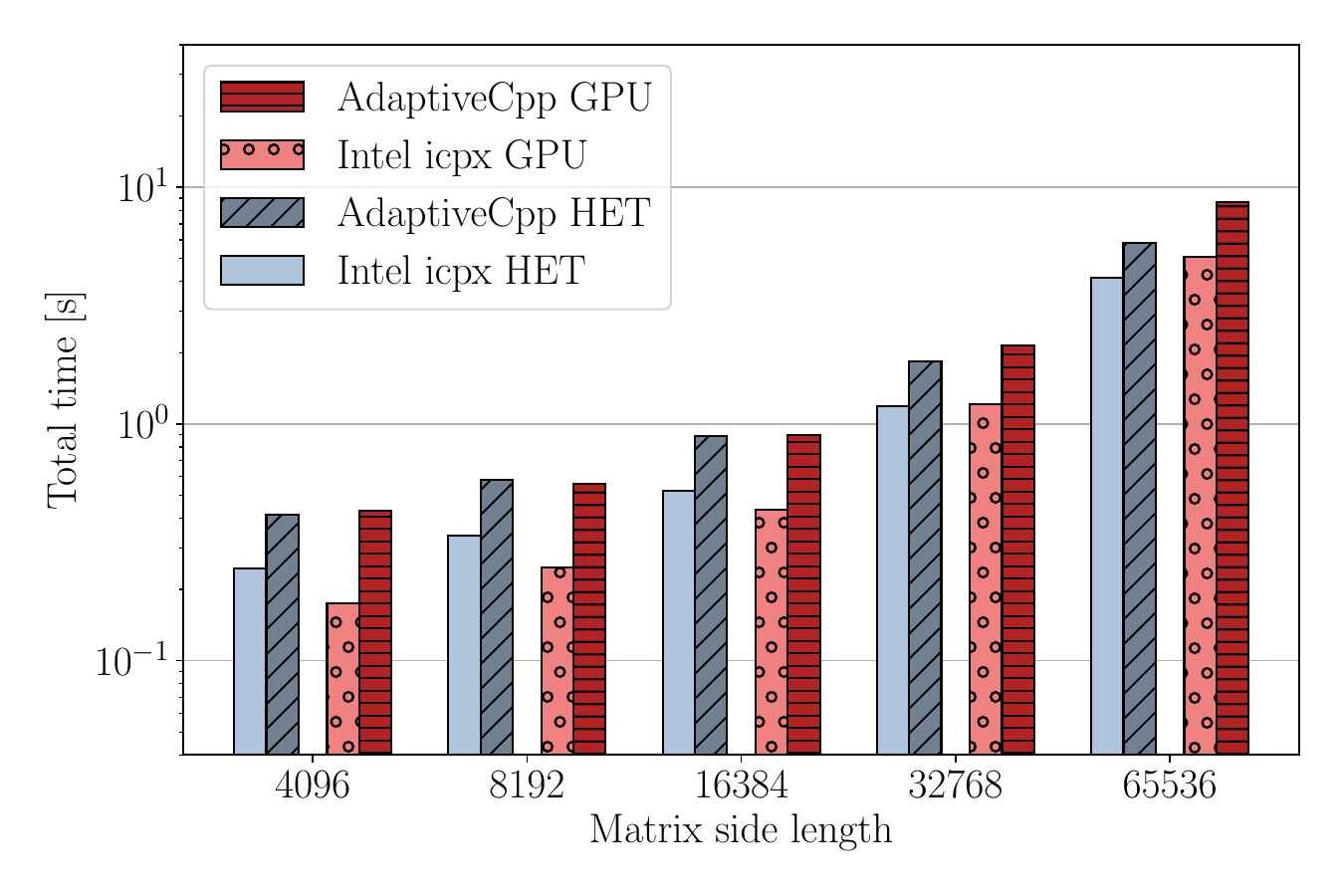}
    \caption{Comparison of the homogeneous and heterogeneous CG algorithm on \textit{System 2} with an AMD MI210 GPU using AdaptiveCpp and the Intel oneAPI DPC++/C++ compiler (icpx). In all cases, icpx results in faster runtimes.}
    \label{fig:cg_icpx_acpp_s2}
\end{figure}

We ran the same experiment for \textit{System 2}, which comprises an AMD MI210 GPU.
The results are presented in \Cref{fig:cg_icpx_acpp_s2}.
Since this system features the same CPU as \textit{System 1} from the previous experiment, we only analyze the GPU-only and heterogeneous performance for this system.
When we examine the GPU-only runtimes, the results show that the Intel icpx compiler results in faster runtimes for all matrix sizes. 
For the largest matrix with $65536 \times 65536$ entries, icpx takes about \num{5.0827952264}\si{\second} whereas AdaptiveCpp needs \num{8.6799486113}\si{\second} in the same scenario.
Thus, icpx results in \num{41.4421046251016}\si{\percent} faster results on the AMD GPU, which is a much larger difference than on the NVIDIA GPU.
In the heterogeneous case, the CG runtime with icpx is reduced to \num{4.1414780687}\si{\second}, which is about \num{28.9455004080423}\si{\percent} faster than the runtime of \num{5.82859367455}\si{\second} observed using AdaptiveCpp.


\subsection{Heterogeneous Cholesky Evaluation}
\label{sec:results_Chol_hetperf}
In this section, our heterogeneous implementation of the Cholesky decomposition is evaluated on \textit{System 1} and \textit{System 2} using AdaptiveCpp. 
As for the CG algorithm, we first determine the optimal values for the parameters that influence the runtime behavior of the Cholesky decomposition.
Next, different workload distributions between the CPU and GPU are compared on both systems.
Finally, we compare the heterogeneous approach to the homogeneous CPU-only and GPU-only implementations.

\subsubsection{Optimal CPU and GPU Configuration for the Cholesky Decomposition}
\label{sec:results_Chol_opt}

Analogous to the CG algorithm, the runtime of the Cholesky decomposition is also influenced by the OpenMP environment (if the CPU is involved in the computation) and the block size of the matrix data structure.
However, in contrast to the CG algorithm, the block size of the matrix data structure not only influences the work-groups of the SYCL kernels but also the Cholesky decomposition itself since we implemented the blocked algorithm. 
Thus, it is crucial to find the optimal value for this parameter.

First, we consider the optimal OpenMP environment. 
Since AVX proved to yield clear performance improvements for the compute-bound Cholesky decomposition, only OpenMP configurations with AVX enabled are discussed here.
As for the CG algorithm, we analyzed the impact of enabling or disabling thread binding and using 48 or 96 threads on the CPU of \textit{System 1} and the largest matrix size with side length $65536$.
We can observe that using 96 threads yields faster results as the runtime is reduced from \num{93.54711}\si{\second} to \num{84.06909}\si{\second}, and enabling binding lowers the variations in this case.
Thus, we choose a configuration that makes use of SMT and enables binding for all further runs on CPUs and AdaptiveCpp.



Regarding the block size, different values that are a power of two have been tested in a similar manner as for the CG algorithm on all devices used in this work.
The results show that a block size of 128 is optimal for all devices except the Intel Arc B580 GPU, where a value of 64 is optimal.



\subsubsection{Optimal Heterogeneous Workload Distribution}
\label{sec:results_Chol_hetperf_dist}

As for the CG algorithm, the optimal workload distribution between the CPU and GPU has to be determined for the heterogeneous Cholesky implementation on all systems.
In this section, we focus on \textit{System 1} and \textit{System 2}.
Remember that the two systems are equipped with identical CPUs.
Thus, the GPU is the only difference when comparing the heterogeneous workload distributions.

\begin{figure}[htb]
    \centering
    \includegraphics[width=\columnwidth]{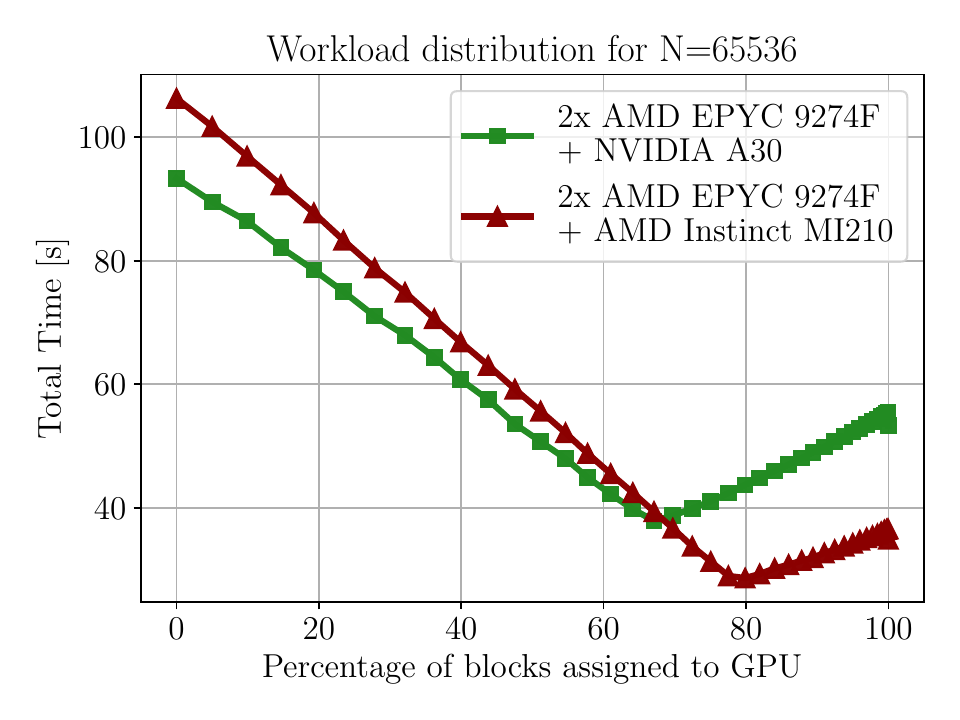}
    \caption{Different workload distributions for the heterogeneous Cholesky decomposition on \textit{System 1} and \textit{System 2}.
    On \textit{System 1}, the distribution is optimal when 67.08\% of the blocks are assigned to the GPU during the matrix multiplication step, whereas this optimum is 79.87\% on \textit{System 2}.}
    \label{fig:chol_het_split_S1_S2}
\end{figure}

\Cref{fig:chol_het_split_S1_S2} shows the runtime of the heterogeneous Cholesky decomposition of a matrix with $65536 \times 65536$ entries for different workload distributions.
As described in \Cref{sec:chol_impl}, the lower triangular matrix is split horizontally between the CPU and the GPU.
The runtime of the Cholesky decomposition is dominated by the block updates using matrix-matrix multiplications.
Due to the triangular structure, each block row holds a different number of blocks that are involved in this step.
Thus, the x-axis of \Cref{fig:chol_het_split_S1_S2} denotes the proportion of blocks that are updated by the GPU in this step and not the proportion of block-rows.
The y-axis shows the runtime of the heterogeneous Cholesky decomposition without memory transfer times.
When analyzing the two minima in the figure, which correspond to the optimal workload distribution between the CPU and the GPU, we can observe that \textit{System 1} reaches the lowest runtime when \num{67.07512842465754}\si{\percent} of the blocks are assigned to the NVIDIA A30 GPU during the update with matrix-matrix multiplications.

For \textit{System 2}, the optimal distribution is achieved when offloading the computation of \num{79.86867049902152}\si{\percent} of the blocks to the AMD MI210 GPU.
These proportions correspond to the lower 42.5\si{\percent} and 55.0\si{\percent} of the block-rows of the triangular matrix, respectively.
Contrary to the CG algorithm in \Cref{sec:results_CG_hetperf_dist}, the AMD GPU now has a higher proportion of the work assigned to it than the NVIDIA GPU, which is what we would expect based on the theoretical performance.
The different behavior compared to the CG algorithm can be explained by the fact that the Cholesky decomposition is dominated by the compute-bound matrix-matrix multiplications.
Thus, the AMD GPU performs closer to its theoretical peak performance here than for the memory-bound CG algorithm.
When considering the runtime where the heterogeneous Cholesky implementation is CPU-bound, we can observe that \textit{System 2} achieves slower runtimes than \textit{System~1}.
However, in contrast to the CG algorithm, the optimal block size is the same on both systems.
Thus, we would expect the same runtime for the CPU-bound case on both systems.
A potential explanation for this behavior is that the memory allocation is performed using the GPU device context in SYCL.
This could enable the underlying SYCL implementation to perform GPU-specific optimizations that resulted in significantly faster memory transfer times between the CPU and GPU in our case.
However, these optimizations might have a negative impact on the CPU performance when optimizing for AMD GPUs.

\subsubsection{Comparison of the Heterogeneous and Homogeneous Cholesky Decomposition Performance}
\label{sec:results_Chol_hetperf_rt}
After the optimal workload distribution analysis in the last section, this section compares the performance of the heterogeneous Cholesky implementation to the corresponding homogeneous implementations that leverage only the CPU or GPU.
The comparison is performed for different matrix sizes.
The optimal heterogeneous workload distributions for the remaining matrix sizes have been determined as before. For the largest three matrices under consideration, the optimal percentages of blocks assigned to the GPU are similar as they range from \num{66.77}\si{\percent} to \num{69.76}\si{\percent} on \textit{System 1} and \num{77.39}\si{\percent} to \num{79.87}\si{\percent} on \textit{System 2}. For smaller matrices, the optimal workload distributions tend towards \num{100}\si{\percent} of the work assigned to the GPU.

\begin{figure}[htb]
    \centering
    \includegraphics[width=\columnwidth]{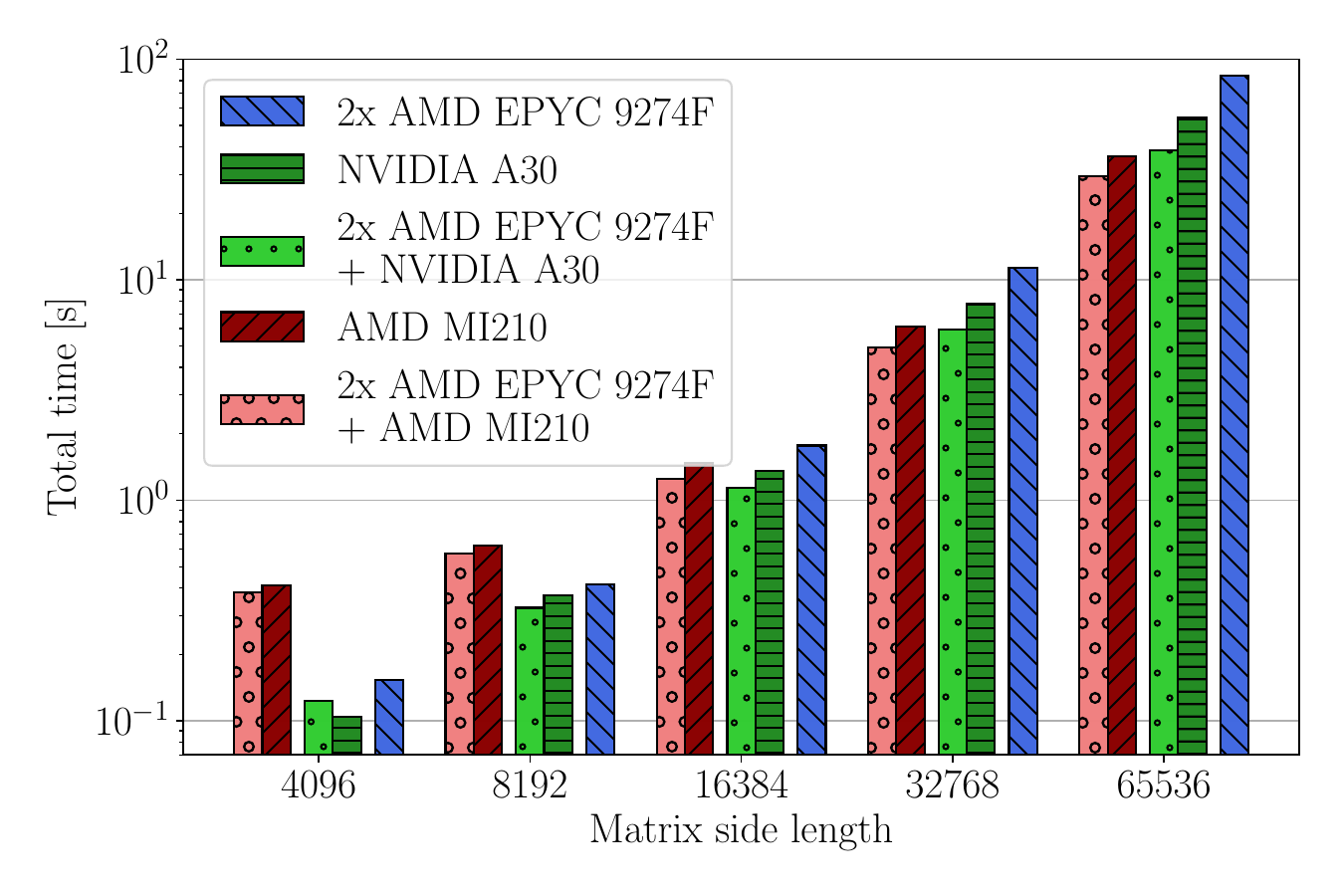}
    \caption{Heterogeneous and homogeneous runtime comparison for the Cholesky decomposition on all hardware of \textit{System 1} and \textit{System 2}. On both systems, the heterogeneous approach results in better performance for large matrices.}
    \label{fig:chol_perf_S1_S2}
\end{figure}

\Cref{fig:chol_perf_S1_S2} shows the results of the comparison on \textit{System 1} and \textit{System 2}.
The x-axis denotes the matrix side length, and the y-axis corresponds to the runtime, including the memory transfer time between the CPU and the GPU. 
When considering the results, we can observe that for both systems, the heterogeneous approach is the fastest method to compute the Cholesky decomposition for the largest three matrices.
For the largest matrix, the CPU-only variant yields the worst performance with \num{84.0919400111}\si{\second} to finish the computation.
Running the homogeneous GPU-only implementation in this scenario takes \num{54.522702870500005}\si{\second} on the NVIDIA A30 GPU and \num{36.3002332649}\si{\second} on the AMD MI210 GPU.
These observations reflect our expectations regarding the theoretical FP64 performance, with the CPU being the least powerful device and the AMD GPU being the most powerful.
However, since the AMD GPU has about \num{4.3461} times the theoretical performance of the NVIDIA GPU, a larger performance difference would be expected. 
The heterogeneous implementation of the Cholesky decomposition has a runtime of \num{38.53193129670001}\si{\second} on \textit{System 1} and \num{29.480133879300002}\si{\second} on \textit{System 2}.
In comparison to the GPU-only runtimes, this is a relative improvement of \num{29.32864794282226}\si{\percent} on \textit{System 1} and \num{18.788031844948495}\si{\percent} on \textit{System 2}.
Thus, the heterogeneous implementation is able to achieve considerable improvements over the homogeneous implementation on both systems.

\subsection{Comparison of AdaptiveCpp and Intel icpx using the Cholesky decomposition}

As for the CG algorithm, we now compare the performance of the Cholesky decomposition using the AdaptiveCpp SYCL implementation and the Intel oneAPI DPC++/C++ compiler icpx.
To allow for a fair comparison between the two compilers, the optimal block sizes and workload distributions for the Cholesky decomposition have been determined for icpx as well.
In contrast to AdaptiveCpp, the CPU configuration for icpx did not have a significant performance impact.
Thus, the default values were chosen.
For AdaptiveCpp, a row-wise parallelization strategy resulted in the best performance for the matrix-matrix step.
However, for icpx, we observed that parallelizing over all result values instead of just the rows resulted in better performance.

\begin{figure}[htb]
    \centering
    \includegraphics[width=\columnwidth]{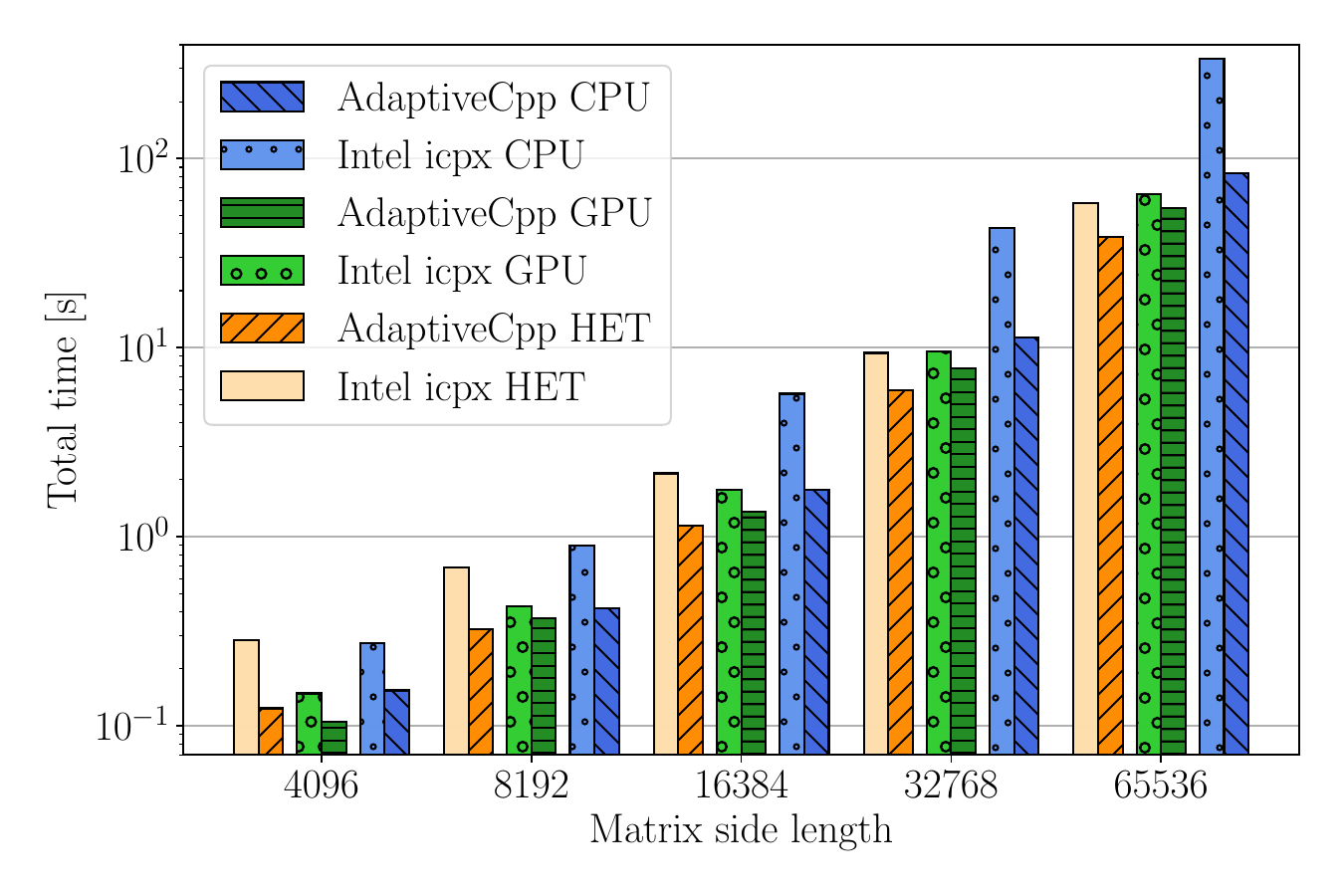}
    \caption{Comparison of the homogeneous and heterogeneous Cholesky decomposition on \textit{System 1} with an NVIDIA A30 GPU using AdaptiveCpp and the Intel oneAPI DPC++/C++ compiler (icpx).
    Since we were not able to leverage vectorization on the CPU with icpx, the CPU runtimes are slower, which also affects the heterogeneous performance of icpx. Nevertheless, AdaptiveCpp also results in faster GPU-only runtimes.}
    \label{fig:chol_icpx_acpp_s1}
\end{figure}

\Cref{fig:chol_icpx_acpp_s1} shows the results of the comparison on \textit{System 1}.
When considering the CPU-only results, we can observe that, in contrast to the results for the CG algorithm in \Cref{sec:results_CG_app_icpx}, icpx takes much longer for the computation than AdaptiveCpp.
For example, when considering the largest matrix, the Cholesky decomposition with icpx takes \num{4.02838133477935} times longer than when using AdaptiveCpp.
This can be explained by the fact that icpx does not make use of vectorization during the matrix-matrix multiplication step of the Cholesky decomposition.
With AdaptiveCpp, we were able to enable vectorization using a simple pragma at the corresponding for-loop. 
However, we were not able to achieve vectorization of this loop with pragmas and icpx.
Thus, using vectorization in our Cholesky implementation with icpx would likely require a complete redesign of our compute kernel for this step, for example, with the usage of SYCL sub-groups.
When we analyze the GPU-only results obtained on the NVIDIA A30 GPU in this system, we observe that AdaptiveCpp is \num{16.1576785052505}\si{\percent} or \num{10.5073462723}\si{\second} faster than icpx, which takes about \num{65.0300491428}\si{\second} for the largest matrix.
When using icpx in the heterogeneous case, it results in slower runtimes than the corresponding AdaptiveCpp results on \textit{System 1}.
This can be explained by the slower CPU and GPU runtimes of icpx.
Overall, icpx can achieve \num{10.5340457849838}\si{\percent} faster runtimes when computing the Cholesky decomposition of the largest matrix heterogeneously instead of homogeneously on the GPU. 
However, the heterogeneous runtime of \num{58.1797539921}\si{\second} is still \num{19.6478226954}\si{\second} slower than the heterogeneous AdaptiveCpp runtime and even \num{3.6570511216}\si{\second} slower than the GPU-only AdaptiveCpp runtime.
For smaller matrices, the heterogeneous Cholesky implementation in combination with the icpx compiler is not able to achieve considerable performance improvements and is slower than the GPU-only version for the smallest three matrices.

\begin{figure}[htb]
    \centering
    \includegraphics[width=\columnwidth]{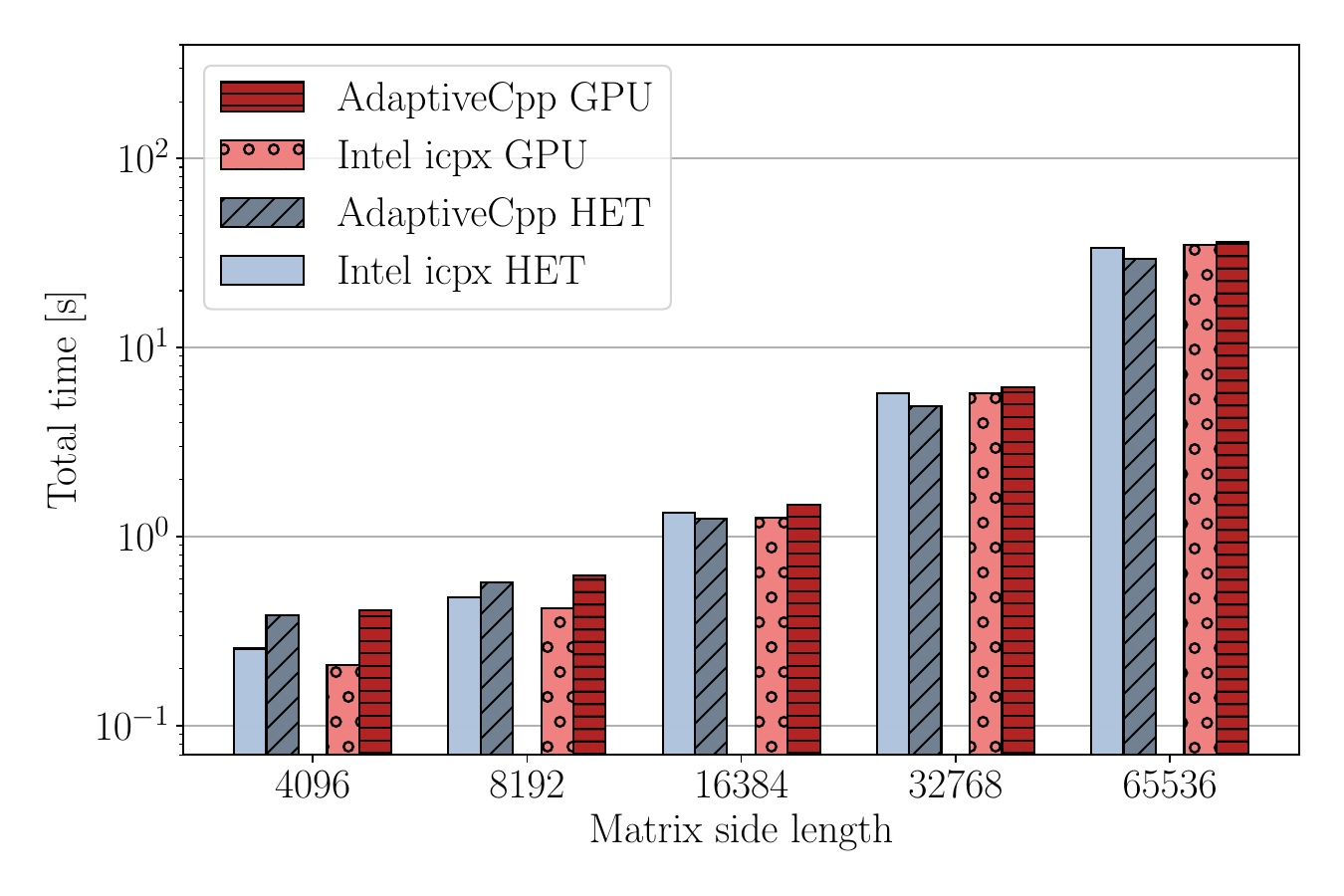}
    \caption{Comparison of the homogeneous and heterogeneous Cholesky decomposition on \textit{System 2} with an AMD MI210 GPU using AdaptiveCpp and the Intel oneAPI DPC++/C++ compiler (icpx). In most cases, icpx results in faster runtimes.
    }
    \label{fig:chol_icpx_acpp_s2}
\end{figure}

We ran the same experiment on \textit{System 2} featuring an AMD MI210 GPU.
The results are presented in \Cref{fig:chol_icpx_acpp_s2}.
Since this system has an identical CPU as \textit{System 1} installed, only the heterogeneous and GPU-only implementations are considered.
When considering the GPU-only case, we can see that, in contrast to the observations on the NVIDIA GPU, icpx results in slightly faster runtimes for the AMD MI210.
For the largest matrix, icpx is about \num{4.17951666929649}\si{\percent} faster than AdaptiveCpp and takes \num{34.7830589646}\si{\second} instead of \num{36.3002332649}\si{\second}.
However, in comparison to the analogous experiment for the CG algorithm in \Cref{sec:results_CG_app_icpx}, the difference between the two SYCL implementations on this AMD GPU is much smaller.
In the heterogeneous case, icpx is only able to achieve very minor or no improvements over the GPU-only Cholesky decomposition.
This can be explained by the slow icpx CPU runtimes for the Cholesky decomposition, limiting large heterogeneous speedups.
As a result, the heterogeneous icpx runtime for the largest matrix is about \num{4.08846461119999}\si{\second} slower than when using AdaptiveCpp.

\begin{figure*}[h]
    \begin{subfigure}{0.49\textwidth}
        \centering
        \includegraphics[width=\textwidth]{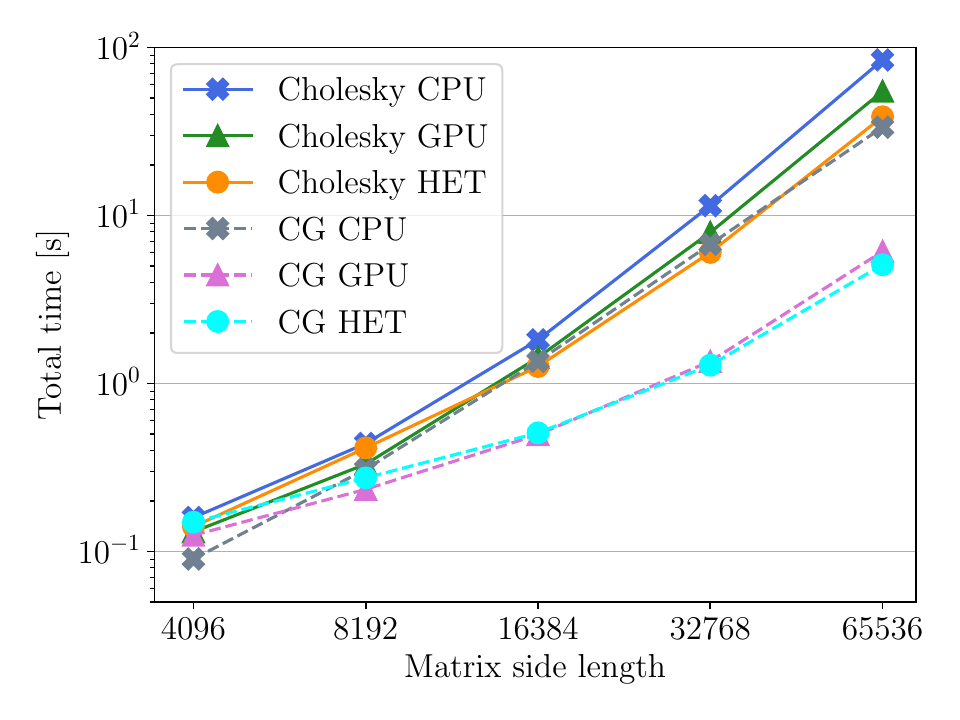}
        \caption{\textit{System 1}}
        \label{fig:alg_comp_S1}
    \end{subfigure}
    \begin{subfigure}{0.49\textwidth}
        \centering
        \includegraphics[width=\textwidth]{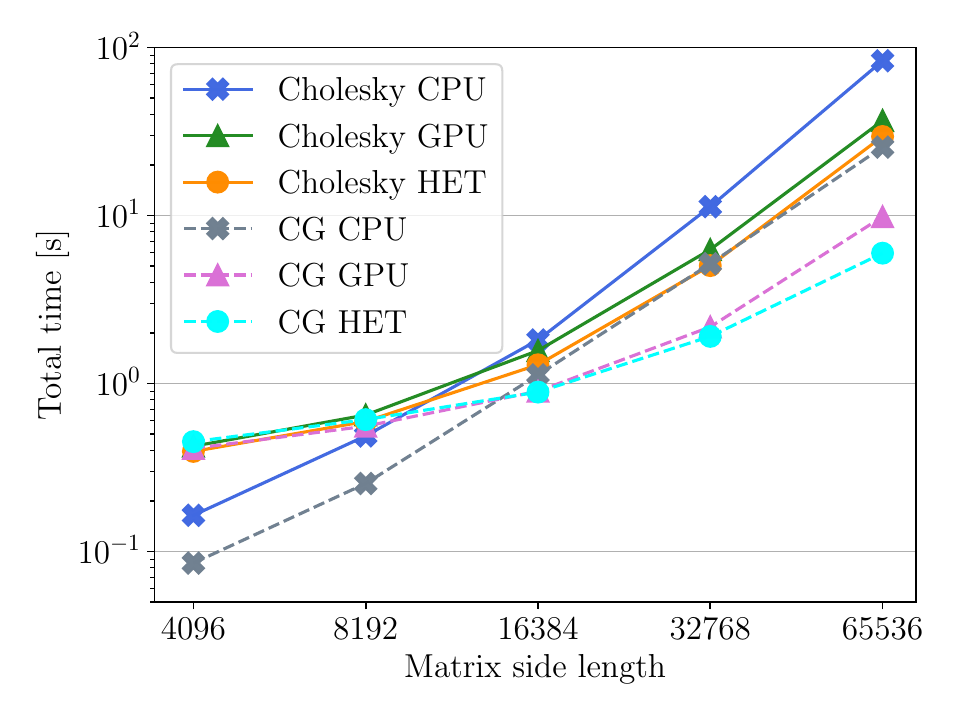}
        \caption{\textit{System 2}}
        \label{fig:alg_comp_S2}
    \end{subfigure}
    \begin{subfigure}{0.49\textwidth}
        \centering
        \includegraphics[width=\textwidth]{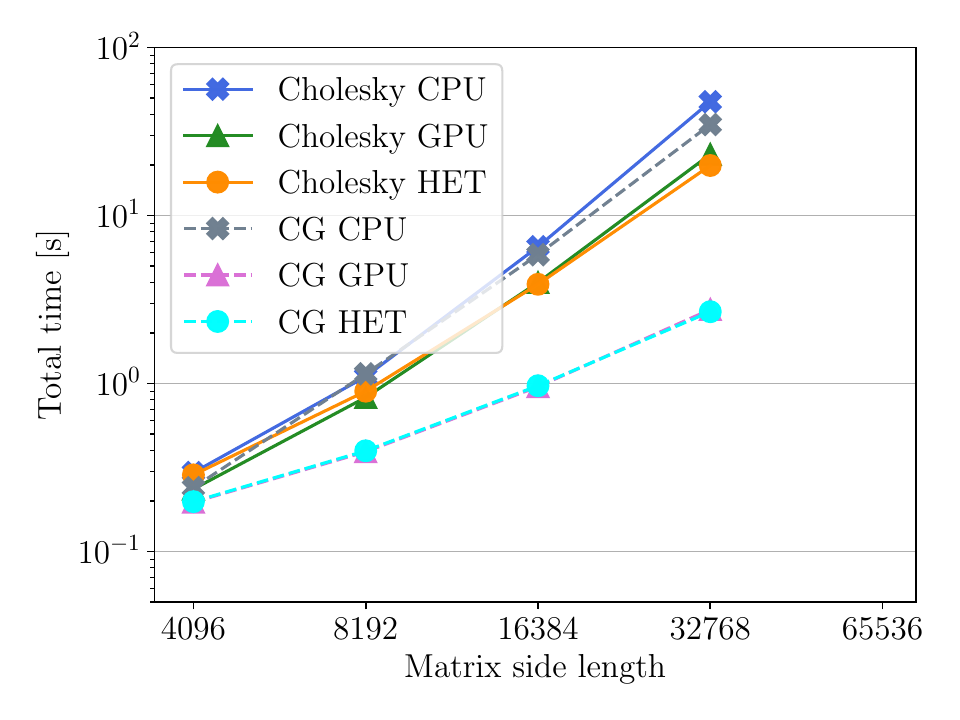}
        \caption{\textit{System 3}}
        \label{fig:alg_comp_S3}
    \end{subfigure}
    \begin{subfigure}{0.49\textwidth}
        \centering
        \includegraphics[width=\textwidth]{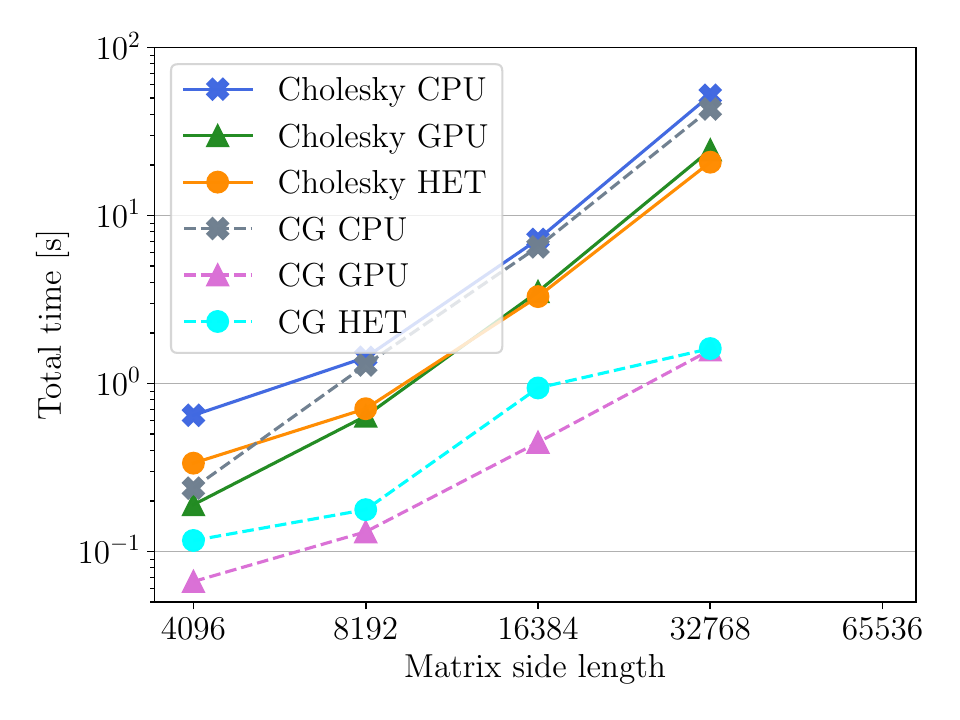}
        \caption{\textit{System 4}}
        \label{fig:alg_comp_S4}
    \end{subfigure}
    \caption{Comparison of the homogeneous and heterogeneous CG algorithm and Cholesky algorithm on all hardware found in the four test systems described in \Cref{tab:systems}.
    In general, the GPU-only and heterogeneous versions of the CG algorithm perform best. However, the CG algorithm generally yields a result with lower precision.}
    \label{fig:alg_comp}
\end{figure*}

\subsection{Comparison of the CG Algorithm and the Cholesky decomposition}
\label{sec:results_CG_Chol_comp}

In this section, we compare the heterogeneous and homogeneous implementations of the CG algorithm and the Cholesky decomposition.
To allow for a fair comparison between the two algorithms, a few changes have to be made to the experimental setup in comparison to the previous sections.
First, the CG runtimes are now measured without the iteration limit to ensure that a result with the desired accuracy of $\epsilon = 10^{-6}$ is obtained in every scenario.
It has to be noted that, due to the specified $\epsilon$-tolerance, the CG algorithm generally yields a less precise result than the Cholesky decomposition.
For the Cholesky decomposition, the solve step involving a forward and back substitution with the triangular Cholesky factor is included in the runtimes to ensure that the same problem is solved by the two algorithms. 
The solve step is not implemented heterogeneously and is performed on the GPU (GPU-only implementation) or the CPU (CPU-only and heterogeneous implementation).
Nevertheless, this step has only a minor impact on the overall runtime.
All measurements are performed using AdaptiveCpp.
Since we were not able to make use of CPU vectorization with icpx, using AdaptiveCpp allows for a fairer comparison between the two algorithms.

\Cref{fig:alg_comp} shows the results of the comparison.
In addition to the two data center-grade systems from the previous experiments, this comparison is extended to two consumer systems featuring an Intel Arc B580 GPU, an NVIDIA RTX 3080 GPU, and identical Intel i9-10980XE CPUs.

First, we consider the results obtained on \textit{System 1} shown in \Cref{fig:alg_comp_S1}.
When solving the linear system involving the largest matrix on the NVIDIA A30 GPU, the CG algorithm can finish the computation \num{8.979187086239463} times faster than the Cholesky solver.
In the heterogeneous case, the CG algorithm results in a \num{7.60100744069275} times faster runtime.
When comparing these findings to the measurements from \textit{System 2}, we can observe that the speedup of the CG algorithm is lower than on \textit{System 1}.
The GPU-only CG implementation on the AMD MI210 is \num{3.7329043514506495} times faster than the corresponding Cholesky implementation for the largest matrix.
With the heterogeneous implementation, a \num{4.949736922073657} times faster runtime is observed in this case.
Furthermore, for small matrices, we can observe that the GPU-only and heterogeneous implementations are not performing well.
This behavior can be explained by higher memory initialization times that were observed on the AMD GPU.
For larger matrices, this effect becomes negligible in relation to the overall runtime.
When focusing on the CPU-only implementation on these two systems, we can observe that the difference between the CG algorithm and the Cholesky solver is smaller. 
Nevertheless, the CG method is still \num{2.514495001807831} times faster on \textit{System 1}  when solving the system involving the largest matrix.

Next, we consider the results on \textit{System 3}, which features an Intel Arc B580 consumer GPU.
Since the largest matrix does not fit into the main memory of this GPU, no measurements can be obtained in this case.
In general, the heterogeneous implementation of the CG algorithm is only able to achieve very minor or no improvements at all. 
In case of the Cholesky decomposition, the heterogeneous implementation achieves a relative performance improvement of up to \num{14.25336748792686}\si{\percent} over the GPU-only implementation.
When comparing the runtime of the CG method with the Cholesky decomposition on the GPU of \textit{System 3}, we can observe that the CG method is \num{8.379789193830687} times faster.
In the heterogeneous case, this factor reduces slightly to \num{7.419920655875478}. 
An interesting observation can be made when focusing on the CPU runtimes. 
Here, the CG method only achieves \num{1.3668528082530114} times faster results than the Cholesky method. 
Furthermore, the CPU-only CG implementation is slower than the GPU-only and heterogeneous implementations of the direct Cholesky solver.
This finding underlines the importance of choosing the right hardware for the right algorithm since the CG method is apparently poorly suited for the CPU in this System.
One possible reason for this may be that \textit{System 3} is using DDR4 memory, which is slower than the DDR5 memory used in \textit{System 1} and \textit{System 2}.

Finally, we examine the results on \textit{System 4} equipped with an NVIDIA RTX 3080.
On this system, the heterogeneous Cholesky implementation achieves  \num{12.575022164392707}\si{\percent} faster runtimes than the NVIDIA RTX 3080 alone.
When considering the CG algorithm, we can observe that the heterogeneous implementation performs equal or worse than the GPU-only implementation on this system.
This observation is similar to \textit{System 3}, which is equipped with an identical CPU.
Since the CG algorithm did not perform well on this CPU, it is also hard to achieve good heterogeneous results. 
Nevertheless, the approach did achieve considerable improvements for the Cholesky decomposition.
When comparing the two algorithms with each other for a matrix with side length $32768$, we can observe that the CG algorithm is \num{15.52699505004494} times faster in the homogeneous GPU-only case. 
When computing heterogeneously on \textit{System 4}, the CG algorithm is \num{12.870149119199919} times faster.
This is a significantly larger speedup compared to the other systems.
For example, on \textit{System 1}, a speedup of just  \num{4.6952551033827055} was achieved by the heterogeneous CG implementation in the same scenario.
This observation could be explained by the fact that the CG algorithm is memory-bound.
Thus, the lack of compute performance of the NVIDIA RTX 3080 GPU in comparison to the NVIDIA A30 GPU plays less of a role for the CG algorithm, but has a huge negative effect for the Cholesky decomposition.

\section{Conclusion}
\label{chap:conl}
In this paper, we presented heterogeneous implementations of the CG algorithm and the Cholesky decomposition that make use of SYCL to simultaneously leverage the CPU and the GPU of the system.
The source code is available on GitHub\footnote{\url{https://github.com/TimThuering/Heterogeneous-Solvers/releases/tag/v1.0.1}}.
We analyzed the performance of the heterogeneous approaches on several systems, which feature GPUs from NVIDIA, AMD, and Intel.
Furthermore, we compared the heterogeneous runtimes with performance measurements of the corresponding homogeneous GPU-only and CPU-only implementations.

\begin{table}[htb!]
	\centering
    \small
	\begin{tabular}{|c||c|c|}
		\hline
						  & \textbf{CG} 				 				&\textbf{Cholesky}						  \\ \hline\hline
		\textbf{System 1} &   \num{12.534446252950126}\si{\percent} (\num{0.6755966110999996}\si{\second})   &    \num{29.32865}\si{\percent}   (\num{15.990771573799996}\si{\second})             \\ \hline
		\textbf{System 2} &   \num{32.84990573605425}\si{\percent} (\num{ 2.85135493675}\si{\second})        &    \num{18.78803}\si{\percent}   (\num{6.820099385599999}\si{\second})              \\ \hline
		\textbf{System 3} &   \num{5.00097}\si{\percent} (\num{0.13669831350000106}\si{\second})             &    \num{14.25337}\si{\percent}   (\num{3.268830302599998}\si{\second})              \\ \hline
		\textbf{System 4} &   \num{0.66740}\si{\percent} (\num{0.01054060170000004}\si{\second})             &    \num{12.57502}\si{\percent}   (\num{3.069703169950003}\si{\second})              \\ \hline
	\end{tabular}
\caption{
	Comparison of the heterogeneous CG and Cholesky performance improvements over the GPU-only implementation for the largest matrix size evaluated on the respective system using AdaptiveCpp. 
}
\label{tab:result_summary}
\end{table}

\Cref{tab:result_summary} summarizes the heterogeneous performance improvements of both algorithms over the corresponding homogeneous GPU-only implementation.
The relative and absolute runtime improvements correspond to the measurements with the largest matrix size considered on the respective system.
In comparison to traditional GPU-only approaches, our heterogeneous CG implementation achieves a performance improvement of up to \num{32.849905736054247}\si{\percent}.
The highest observed value for the Cholesky decomposition corresponds to  \num{29.32865}\si{\percent}.
Furthermore, the same heterogeneous SYCL code for the Cholesky decomposition is at least \num{12}\si{\percent} faster than the GPU-only implementation across all four systems for the respective largest matrices.
A comparison of two popular SYCL implementations, AdaptiveCpp and the Intel oneAPI DPC++/C++ compiler icpx, showed that AdaptiveCpp was able to achieve faster CPU and NVIDIA GPU runtimes for the Cholesky decomposition.
Meanwhile, the Intel compiler resulted in faster runtimes for the CG method.
On the AMD GPU, icpx resulted in faster runtimes for both algorithms.

Future work could explore how the heterogeneous performance differs when using the native language for the respective hardware instead of SYCL. 
This work compared the heterogeneous and homogeneous implementations regarding their runtime.
Future work could investigate how the energy consumption differs between the different algorithms and on different hardware.
Furthermore, the heterogeneous solvers could be extended using mixed-precision.

\subsection*{AI Use Disclosure}

Generative AI tools, including Grammarly~\cite{grammarly}, DeepL~\cite{deepl}, and ChatGPT~\cite{chatgpt}, were employed to enhance the clarity, grammar, and overall coherence of the manuscript. All technical content, data analyses, and research findings were conceived and developed independently by the authors. AI-assisted outputs were carefully reviewed, verified, and edited by the authors to ensure factual accuracy, interpretive rigor, and scholarly integrity. The final manuscript reflects the authors’ original intellectual contributions and analytical work.

\bibliographystyle{ACM-Reference-Format}
\bibliography{references}

\end{document}